\documentclass{style/nseJournal}

\usepackage{wrapfig}
\usepackage{ulem}
\usepackage{amsfonts}
\usepackage{amssymb}
\usepackage{multirow}
\newcommand{\avgA}{$\langle A \rangle \:$}
\begin{document}

\title{Machine-Learning enabled analysis of ELM filament dynamics in KSTAR} %title of paper

% Use the \addAuthor macro to add authors in the order they should appear. The second argument corresponds to
% the affiliation declared below.
% The corresponding author should be wrapped in \correspondingAuthor
\addAuthor{\correspondingAuthor{Cooper Jacobus}}{a}
% The corresponding author's email can be specified using \correspondingEmail
\correspondingEmail{cjacobus@berkeley.edu}
\addAuthor{Minjun J. Choi}{b}
\addAuthor{Ralph Kube}{c}

% Affiliations can be added in the order they should appear. For breaks in addresses, use either \\ or \tabularnewline
\addAffiliation{a}{University of California, Berkeley, CA 94720, USA}
\addAffiliation{b}{Korea Institute of Fusion Energy, Daejeon 34133, Republic of Korea}
\addAffiliation{c}{Princeton Plasma Physics Laboratory, NJ 08540, USA}

% Add keywords to appear in Abstract in the order they should appear
\addKeyword{Machine Learning}
\addKeyword{Convolutional Neural Networks}
\addKeyword{Edge Localized Mode}
\addKeyword{Electron Cyclotron Emission Imaging}

\titlePage

\begin{abstract}
The emergence and dynamics of filamentary structures associated with edge-localized modes (ELMs) inside tokamak plasmas during high-confinement mode 
is regularly studied using Electron Cyclotron Emission Imaging (ECEI) diagnostic systems. ECEI allows to infer electron temperature variations, often
across a poloidal cross-section. Previously, detailed analyses of filamentary dynamics and classification of the precursors to ELM
crashes have been done manually. We present a machine-learning-based model, capable of automatically identifying the position, spatial extend, and
amplitude of ELM filaments. The model is a deep convolutional neural network that has been trained and optimized on an extensive set of manually labeled
ECEI data from the KSTAR tokamak. Once trained, the model achieves a $93.7\%$ precision and allows to robustly identify plasma filaments in unseen ECEI data.
The trained model is used to characterize ELM filament dynamics in a single H-mode plasma shot. We identify quasi-periodic oscillations of the filaments size,
total heat content, and radial velocity. The detailed dynamics of these quantities appear strongly correlated with each other and appear qualitatively
different during the pre-crash and ELM crash phases.
\end{abstract}

\section{Introduction}

High confinement mode (H-mode) plasmas are characterized by a steep density gradient in the edge,
the so-called pedestal region, of the confined plasma. Access to H-mode is achieved by plasma heating and once a critical heating
threshold is exceeded the pedestal typically forms within a few milliseconds. Associated
with this rapid transition are strong, sheared electric drifts, localized to the pedestal region,
as well as a local suppression of local turbulent fluctuations \cite{moyer-1995, wagner-2007, leonard-2014}.
These sheared electric drifts form a transport barrier which inhibits radial transport. As a result,
the energy confinement properties of H-mode plasmas are superior to those in low-confinement mode and render
this mode of operation an attractive baseline scenario for ITER \cite{iter-technicalbasis, kim-2018}.
Edge localized modes (ELMs) are a ubiquitous feature of H-mode plasmas. They cause an intermittent
relaxation of the edge pressure gradient, caused by so-called ELM crashes. During these violent events,
plasma ejects from the confined region onto material surfaces of the vacuum vessel. Upon contact with
the hot plasma, material surfaces erode and sputtered wall atoms contaminate the confined plasma.
These phenomena are at the core of complex research activities in the fusion community. 

%Regulating impurities in magnetically confined plasmas, which includes both heavy wall impurities and fusion ash for future burning plasmas, is critical for their stable confinement \cite{angioni-2021}. Furthermore dictate transient heat loads from ELM cycles the desired properties of plasma facing components for magnetic fusion devices \cite{brezinsek-2017}.

There are various types of ELMs, identified based on periodicity and dependence on heating power. These types include
violent, low-frequency Type-I ELMs, or more intermittent, burst-like Type-III ELMs, as well as a mixed regime
\cite{hill-1997, lee-2001, ahn-2012}. While two-fold stability analysis of peeling-ballooning modes explains linear
stability of ELMs \cite{snyder-2005, ham-2020}, the actual ELM crash is a nonlinear magnetohydrodynamic phenomenon
and an area of active research interest \cite{snyder-2011, leonard-2014, oh-2018, kim-2020, ham-2020, Kim_2020}. 

At the KSTAR Tokamak, a two-dimensional (2D) Electron Cyclotron Emission imaging diagnostic \cite{yun-2010, yun-2014} has been 
used to investigate the nonlinear dynamics of ELMs, including their explosive growth, saturation, and crash \cite{yun-2012, park-2019}.
In the 2D ECE images, the ELM mode appears as a filamentary structure of the normalized temperature fluctuation near the pedestal top.
The evolution of these ELM filaments is divided into three phases. During an initial growth phase, 
the amplitude of the mode structure increases while the structure itself rotates counterclockwise
through the ECEI field of view. Once saturated, the ELM filament amplitude appears to stagnate while the filaments
into finger-like structures that extend from the pedestal into the open field-line region.
A similar phenomenology is reported from other tokamaks \cite{boedo-2005, terry-2007, maqueda-2009}.
In order to investigate physical mechanisms that drive ELM filament dynamics and lead up to ELM crashes
it is desirable to robustly estimate properties of the structures visible in the ECEI data. This includes 
for example the number of visible filaments, their amplitude, their spatial extent, and their velocities. 
With this information at hand, one can investigate all three phases, initial growth, saturation, as well
as the ELM crash itself. This work introduces a machine-learning based model that allows automatic
frame-by-frame detection of filamentary structures associated with ELMs. We do not distinguish between the
turbulent inter-ELM phase and the non-linear evolution of the actual ELM crash. For a recent experimental
review, the interested reader is referred to \cite{diallo-2021}.

%\cite{kirk-2004} Langmuir probe measurements at the MAST tokamak, in combination with $D_\alpha$ and
%high-speed video images of the entire plasma column suggest large radial heat flux events,
%mediated by ELM filaments. Since then they have been observed on most tokamaks, and a recent review
%by Ham et ak. gives a detailed review over experimental findings \cite{ham-2020}.
%Diagnosed using doppler backscattering in the Globus-M tokamak \cite{bulanin-2019}.

Given its ability to learn complex non-linear relations from large amounts of data, machine learning is applied
to various tasks in fusion energy research. A major area of concern for developing magnetic confinement based
fusion reactors is the safe shutdown of pulses before explosive disruptions exert intolerable stresses on
plasma facing components. Besides disruption detection, it is a major goal to integrate machine-learning based detection
algorithms into real-time plasma control systems in order to enable robust plasma control solutions
\cite{kates-harbeck-2019, Sias-2019, Pau-2019, Churchill-2020, Fu-2020, Guo-2020, Rea-2020, Boyer-2021, Barr-2021, Hu-2021, Degrave-2022, aymerich-2022}.
Reduced, or surrogate models are another successful use case of machine learning models in fusion energy sciences.
Such models allow to instantaneously infer predictions of high-fidelity models for a given set of inputs. Surrogate
models are for example relevant for complex optimization loops and similar integrated data analysis tasks. Post-shot analysis of plasma pulses for example
can be performed orders of magnitude faster using surrogate models while retaining high accuracy
\cite{Joung-2019, vdPlassche-2020, Ho-2021, kaptanoglu-2021, Li-2021, Dong-2021, Merlo-2021}. 

Convolutional neural network based models have been used to explore connections between bursting behaviour
of the plasma boundary and solitary perturbations \cite{lee-2021}. Filamentary structures, sometimes called blobs
when discussing scrape-off layer plasmas, present a ellipsoid footprint when viewed in the
poloidal - radial plane \cite{zweben-2002}. Automatic detection of such structures is often been performed using blob detection,
which identifies coherent structures of a minimum size that exceed a heuristically determined amplitude threshold 
\cite{kube-2013, zweben-2015, decristoforo-2020}. Other work approaches blob detection using methods based on cross-correlation
of spatial channels \cite{agostini-2011} or spatial displacement estimation \cite{lampert-2021-rsi}. Only recently have machine learning
methods been used to identify the extent of filament
structures in fusion plasmas \cite{imre-2019}. Machine-learning based models learn detection criteria based on labeled
training data. That is, they don't require a-priori heuristically tuned parameters for structure detection but replace
them with information encoded in expert-labeled training data. The approach presented here is of the same spirit and
to the best knowledge of the authors a first attempt to use a machine-learning-based algorithm to investigate ELM filament dynamics.

The remainder of the article is structured as follows. In section \ref{sec:method} we introduce the used deep convolutional
neural network architecture and describe its application for ECEI data. We also describe the developed training data set
and the performance of the trained network on training and validation data sets. In section \ref{sec:analysis} we use
the trained model to identify filamentary structures in a H-Mode plasma shot and explore their dynamical properties.
We discuss the results and set them into context with other recent work in \ref{sec:discussion}. A conclusion and 
directions for future work are given in \ref{sec:conclusion}

%%%%%%%%%%%%%%%%%%%%%%%%%%%%%%%%%%%%%%%%%%%%%%%%%%%%%%%%%%%%%%%%%%%%%%%%%%%%%%%%%%%%%%%%%%%%%%%%%%%%%%%%%%%%%%%%%%%%%%%%%%%%%%%%%%%%%%%

\section{Machine learning detection of ELM filaments in ECEI data}
\label{sec:method}

Cyclotron radiation emission from free electrons in magnetically confined plasmas are routinely measured for diagnostic purposes 
\cite{yun-2010, hutchinson-book}. At KSTAR, 2D ECEI systems are used to measure electron temperature fluctuations
$\delta T_\mathrm{e} / \langle T_\mathrm{e} \rangle$ over a two-dimensional field of view aligned in the poloidal cross-section 
on a microsecond timescale. \cite{yun-2010, yun-2014}. Each system covers rectangular cross-sections of the plasma with 24 by 8 
pixels in the vertical and radial direction respectively, featuring aspatial resolution $2\mathrm{cm}$ and a sampling time of 
$2 \mu \mathrm{s}$. In practice, the low-field side field-of-view extends about $40\, \mathrm{cm}$ vertically and about 
$10\, \mathrm{cm}$ radially which readily allows to visualize filamentary structures associated with ELMs.

\begin{figure}[htb]%[!htbp]
  \centering
  \includegraphics[width=0.9\textwidth]{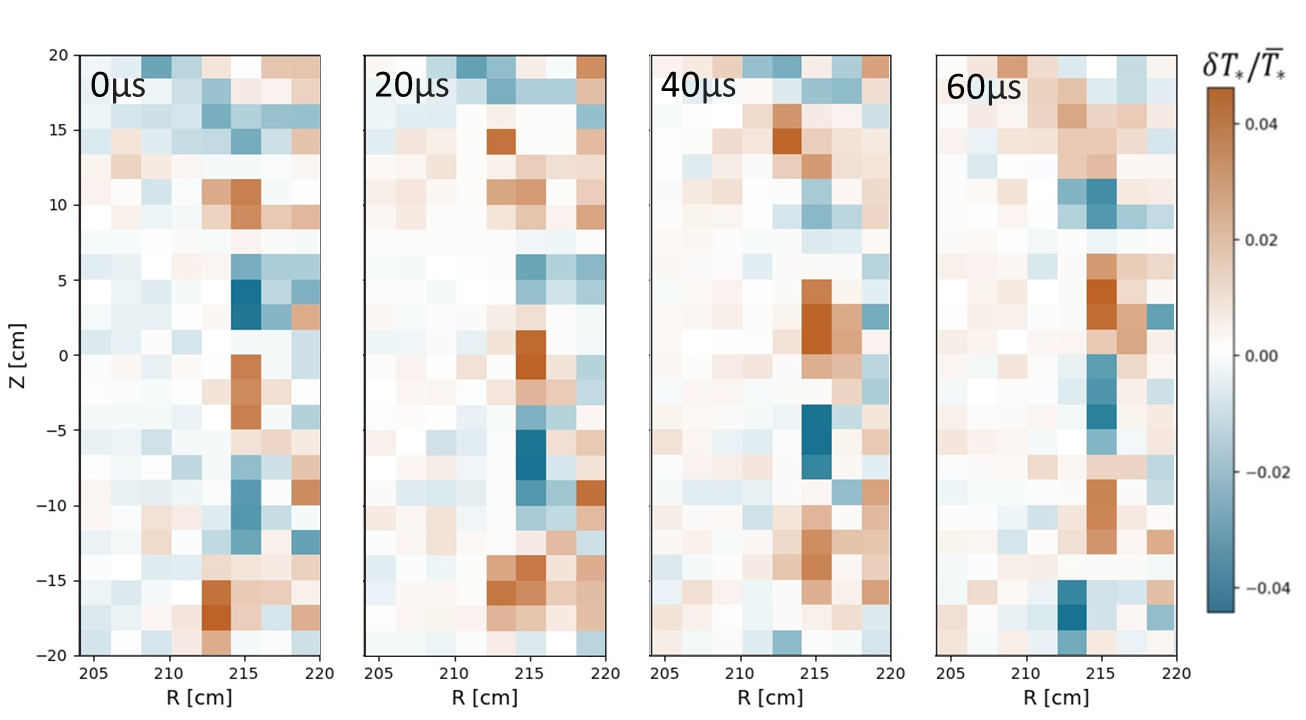}
  \caption{A sequence of ECE images taken during the initial growth phase of an ELM. Humans readily identify 
  a modal structure with peaks and troughs which rotates counter-clockwise in this image sequence.}
  \label{fig:ECEI_Sample}
\end{figure}

Normalizing to vanishing mean, the emission intensity sampled by the ECE diagnostic can be interpreted as a temperature fluctuation $\delta T / \langle T \rangle$.
The temperature corresponds to the radiation temperature and it is usually justified to associate it with the electron temperature 
\cite{yun-2010, hutchinson-book}. All ECEi data used in this contribution is normalized this way, while pixels with low signal-to-noise
ratio are replaced by a linear interpolant. The data was furthermore subject to a bandpass filter of $5-9 \mathrm{kHz}$, 
the range being estimated as to capture the dominant fluctuations in between ELM crashes shown in Fig.~\ref{fig:22289_overview}.
Figure \ref{fig:ECEI_Sample} shows a sequence of bandpass filtered and normalized ECEI images that capture a mode structure
associated with an ELM. To estimate the dynamics present in this sequence of images we can visually identify individual peaks and troughs
in each image. Asserting that peaks (or troughs) in subsequent images with maximal overlap correspond to the same part of the observed mode
structure, we estimate a rotation velocity of approximately $2\, \mathrm{cm}$ per $20 \mu \mathrm{s}$ frame, that is, the modal structure
rotates at roughly $1 \mathrm{km}/\mathrm{s}$. A computer vision algorithm can automatically perform this calculation. For this, a
detector should identify individual peaks in a single frame. Then a tracking algorithm can be used to track these peaks across
a squence of images.

In this contribution, we use a Scaled-YOLOv4 (You Only Look Once) model \cite{redmon-2015, bochlovskiy-2020, yao-2020} to identify 
ELM filaments in ECEI data. YOLO is a widely used class of object detection models and the Scaled-YOLOv4 model modifies 
the original YOLO architecture in order to be applicable to smaller images. Object detection models, such as YOLO, perform multi-class
object detection on image data by outputting bounding box dimensions as well as a probability of belonging to a given class
for each identified object. The YOLO model is based on a convolutional neural network (CNN) architecture, which applies a sequence
of learned filters, each followed by non-linear activation functions, to input data.  The size and the stride as well as the number
of output channels of the convolution filters are fixed while the convolution matrix coefficients are trainable parameters.
Besides its convolutional architecture, design choices in the YOLO architecture are made in order to optimize inference speed,
which makes it one of the fastest image detection algorithms available \cite{bochlovskiy-2020}. YOLO models also perform
exceptionally well, for example it has been found that they detect approximately three times fewer false positives than 
alternative architectures \cite{Jiao}.

\begin{figure}[htb]
  \centering
  \includegraphics[width=1.0\textwidth]{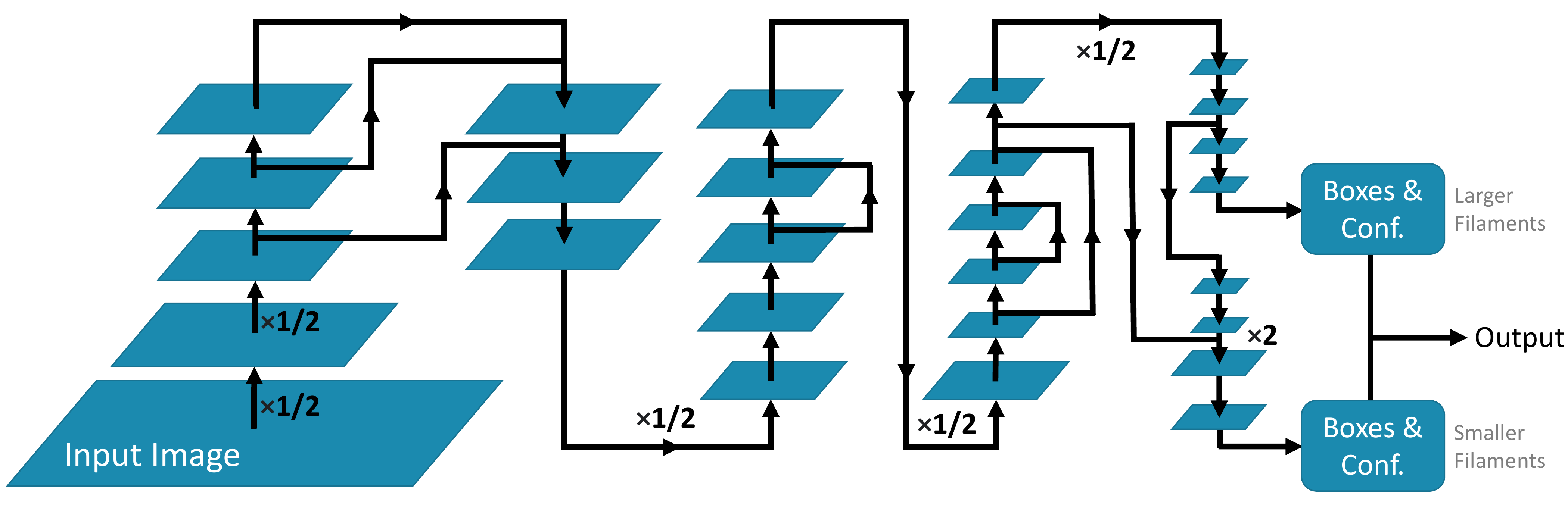}
  \caption{Illustration of the convolutional architecture implemented by the used YOLOv4 model. Blue rectangles denote data
           at a given layer and their size is proportional to the matrix dimensions. Black arrows denote connections between layers, and
           are annotated by scale factors in the case where a convolution changes the layer size. There are two outputs layers, each of the
           giving a set of bounding boxes as well as a confidence score for detected instances of object classes in the input image.}
  \label{fig:Architecture}
\end{figure}

Figure \ref{fig:Architecture} illustrates the architecture of the used Scaled YOLOv4 network. The input is a 192-by-64 pixel image
with 3 channels. To transform the normalized $24 \times 8$ ECEI data frames to this size they are up-sampled using cubic interpolation
and then a colormap is applied.  The first convolution applies a set of $32$ filters, each of size $3 \times 3 \times 3$ using a stride of 
2 and outputs $32$ $92 \times 32$ matrices, reducing the input resolution by a factor of 2. The values of these filters are the 
trainable parameters of the model. The output of the first layer consists of convolutions of the input image with each one of these 32
filters. The resulting stack of matrices is then used as input to the next layer. These 32 matrices are 'feature maps' which represent the
qualitative properties of a local neighborhood of pixels \cite{feature-1, feature-2}. Layers of these convolutions are applied to produce more
specific feature maps until it is able to recognize complex features like filaments \cite{CNNs}. While connections are mostly between
successive layers, YOLO also uses a number of residual connections, where the outputs of non-consecutive layers are concatenated along the
channel dimension. The outputs of the YOLO network are two sets of bounding boxes with associated confidence scores. The first set of outputs
is used to identify larger structures. Then, continuing processing through some more up-convolution layers, another output is generated that
aims to identify smaller structures. Reference \cite{bochlovskiy-2020} describes the architecture in detail. For the work presented in this
article we used the scaled YOLO-v4 implementation provided here \footnote{\url{https://github.com/AlexeyAB/darknet}}.

\begin{figure}[h!tb]
  \centering
  \includegraphics[width=1.0\textwidth]{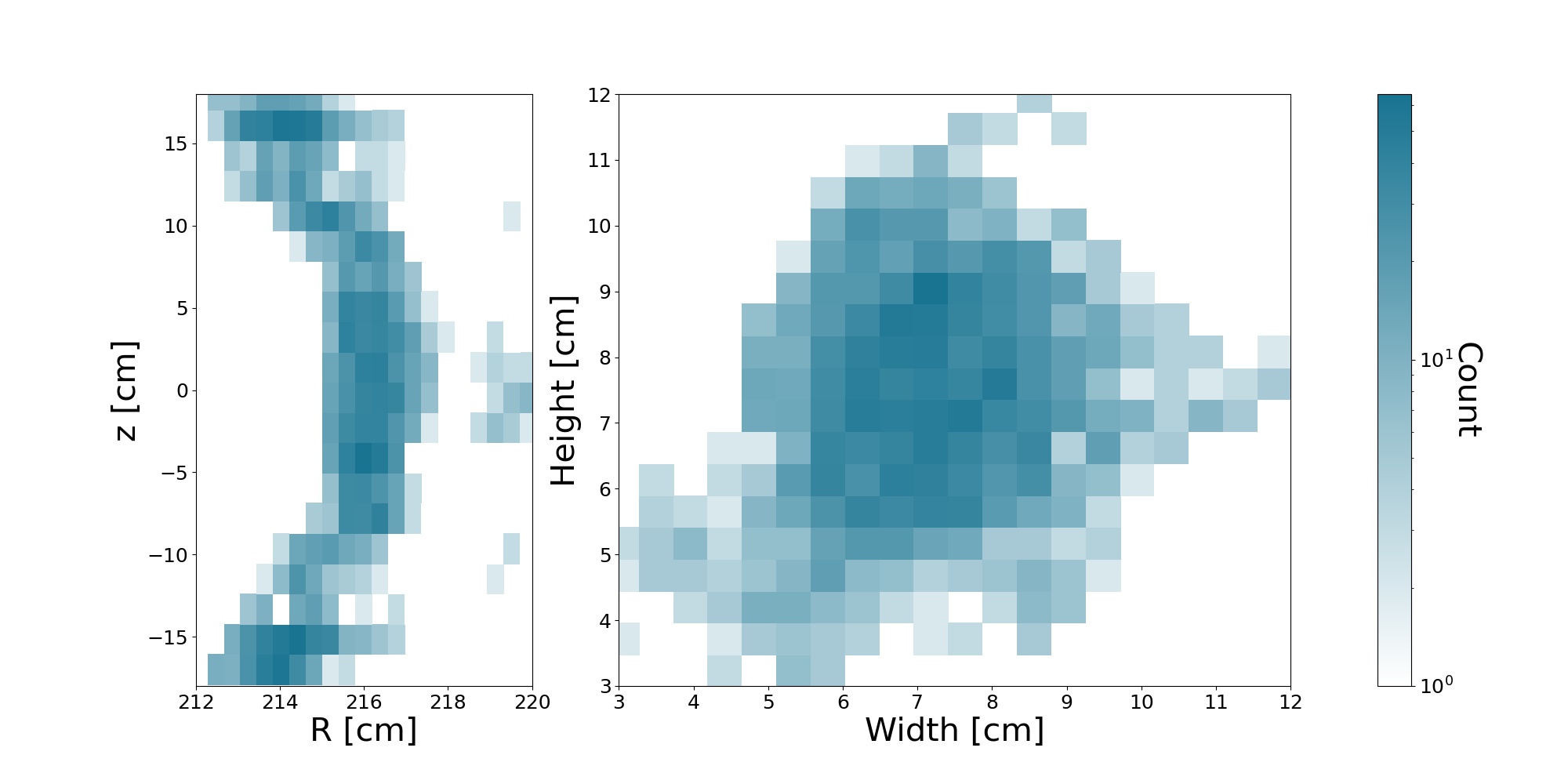}
  \caption{Histograms of the anchor and dimensions of the bounding boxes prescribed in the manually labelled training data.}
  \label{fig:label_dist}
\end{figure}

YOLO is trained in a supervised fashion and requires labeled training data. We manually created a training set of ELM filaments by
randomly selecting 1000 ECEI data frames from a single KSTAR shot and fit tight bounding boxes around ELM filaments that appear in these frames. 
A bounding box is described by four values: an anchor point in the 2d ECEI view, a width, and a height. We only label
hot filaments which contain excess heat and therefore all bounding boxes are assigned the same class. Figure \ref{fig:label_dist} 
shows the distribution of bounding boxes in the training set. The anchors appear mostly on a singular flux surfaces with some detections also 
at $R \approx 220 \mathrm{cm}$ and $Z\approx 0\,\mathrm{cm}$. The width and height of the bounding boxes distribution in the manually
labeled data are approximately $7 \pm 3 \mathrm{cm}$. There are only a few outliers in the bounding box size distributions, either centered at smaller sizes or tall filaments with a width of approximately $10\, \mathrm{cm}$ and a height of approximately $7 \mathrm{cm}$.

The scope of this paper is intended to be an analysis of a single KSTAR shot, all frames included in the training data are from the range 2.6-2.8 seconds.
The frames used for labeling were drawn from a uniform distribution. So they represent the temporal distribution of pre-growth,
mid-growth, saturated, and mid-crash phases. On average, there is about $ 200 \mathrm{\mu s}$ between any two frames, long enough that each
there is negligible correlation between any two frames of the training set.

\begin{figure}[h!]
  \centering
  \includegraphics[width=0.7\textwidth]{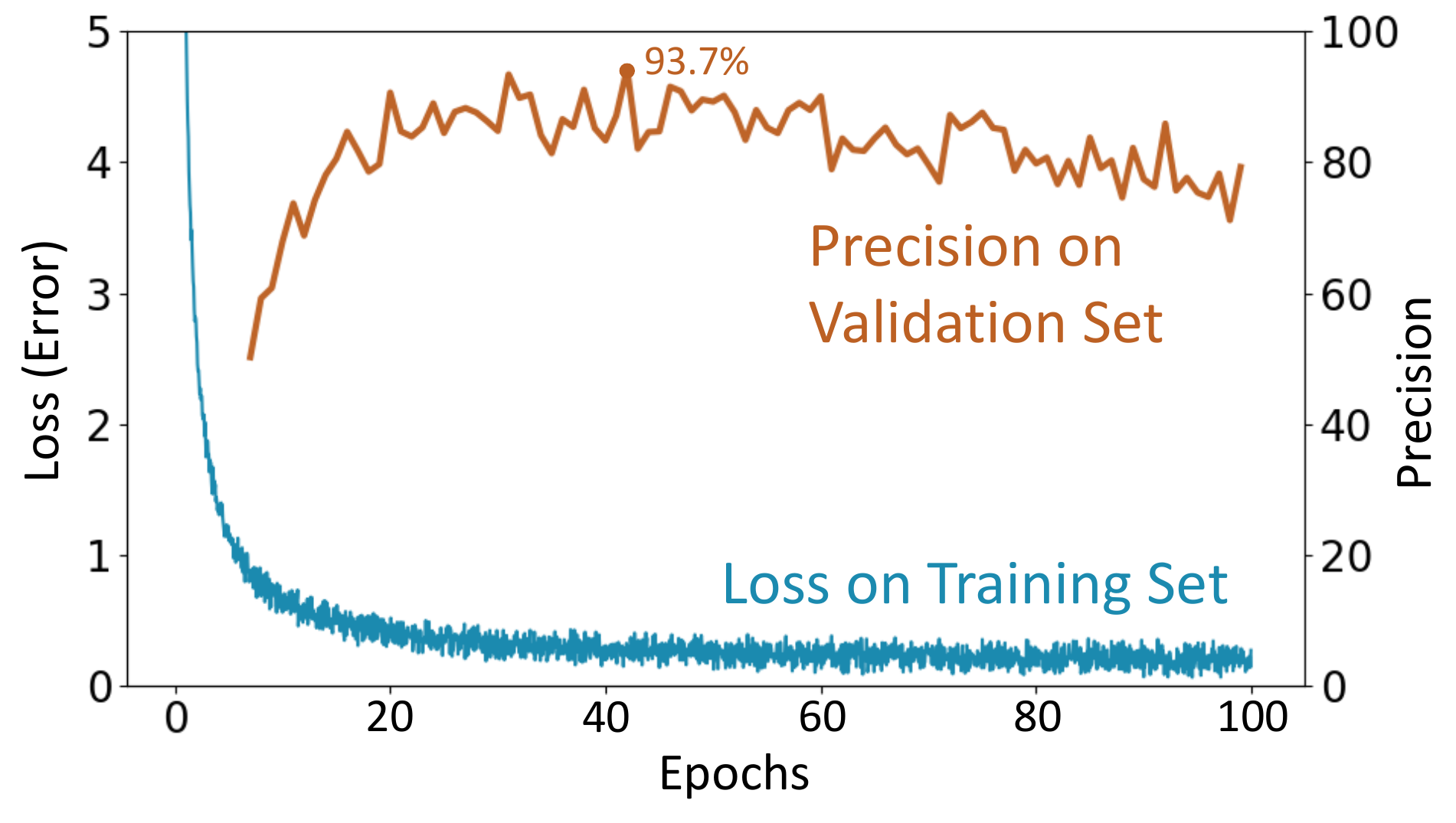}
  \caption{Calculated values of YOLO loss on the training set and precision on the validation set during training.}
  \label{fig:training}
\end{figure}

The task of the trained model is to output tightly fitting bounding boxes around ELM filament instances in unseen data. 
YOLO uses a special loss function which is given by a weighted sum of multiple terms. These include a binary cross-entropy term
for correct instance labeling as well as a term describing intersection-over-union (IoU) misalignment of the bounding boxes. 
Intersection over union is a commonly used quantity in object identification and localization \cite{bochlovskiy-2020}. Given 
two boxes, IoU is calculated as the ratio over the intersection of the boxes to the union. For training, we randomly split the
manually labeled data into a training set and a validation set, 
consisting respectively of 800 and 200 frames with their associated bounding boxes. The model parameters are optimized by minimizing the
YOLO training loss function calculated over the training set for 100 epochs. We used a batch size of 64 randomly shuffled images and 
the Adam optimizer \cite{Kingma-2014} with a learning rate of 0.00261, as recommended in the documentation of the used YOLOv4 implementation
\footnote{\url{https://github.com/AlexeyAB/darknet}}. 

Performance metrics of the model can readily be calculated from its IoU predictions. To aid in consuming the IoU information, it is common to
identify a detection where $IoU \geq 0.5$ as a true positive. Similarly, a false positive detection is an instance where $IoU < 0.5$.
A false negative detection is a case where the model identifies an object instance in an image where it is not present. And finally,
a true negative detection is a case where the model does not identify object instances and they are not present in the image.
During training, the performance of the model is monitored by tracking the value of the loss function calculated over the training 
set and the precision, the ratio between true positives to the sum of true and false positives, calculated on the validation set. 
Figure \ref{fig:training} shows the training loss and validation precision of our model during training. Initially, the training
loss decreases sharply and flattens after about 40 iterations. At the same time, the precision on the validation set increases
from about 50 to over $90\%$ from epoch 10 to 40. 

After 42 epochs of training, the model achieves a $93.7\%$ precision on the validation set. That is, it correctly captures the bounding
box of the majority of the validation data. It also achieves a $85.2\%$ recall on epoch 42, the ratio of true positives to the sum of true
positives and false negatives. In other words, almost all detected filaments are indeed filaments present in the validation set.
Training for more than 42 epochs we observe that the model begins to over-fit on the training set. On average, our model achieves $67.9\%$ IoU,
which highlights that there is still notable disagreement between the two. This stems however mostly from misalignment of the bounding boxes and not from 
mis-detection of filaments. The model state reached in epoch 42 is used for inference.

\begin{figure}[h!tb]
  \centering
  \includegraphics[width=1.0\textwidth]{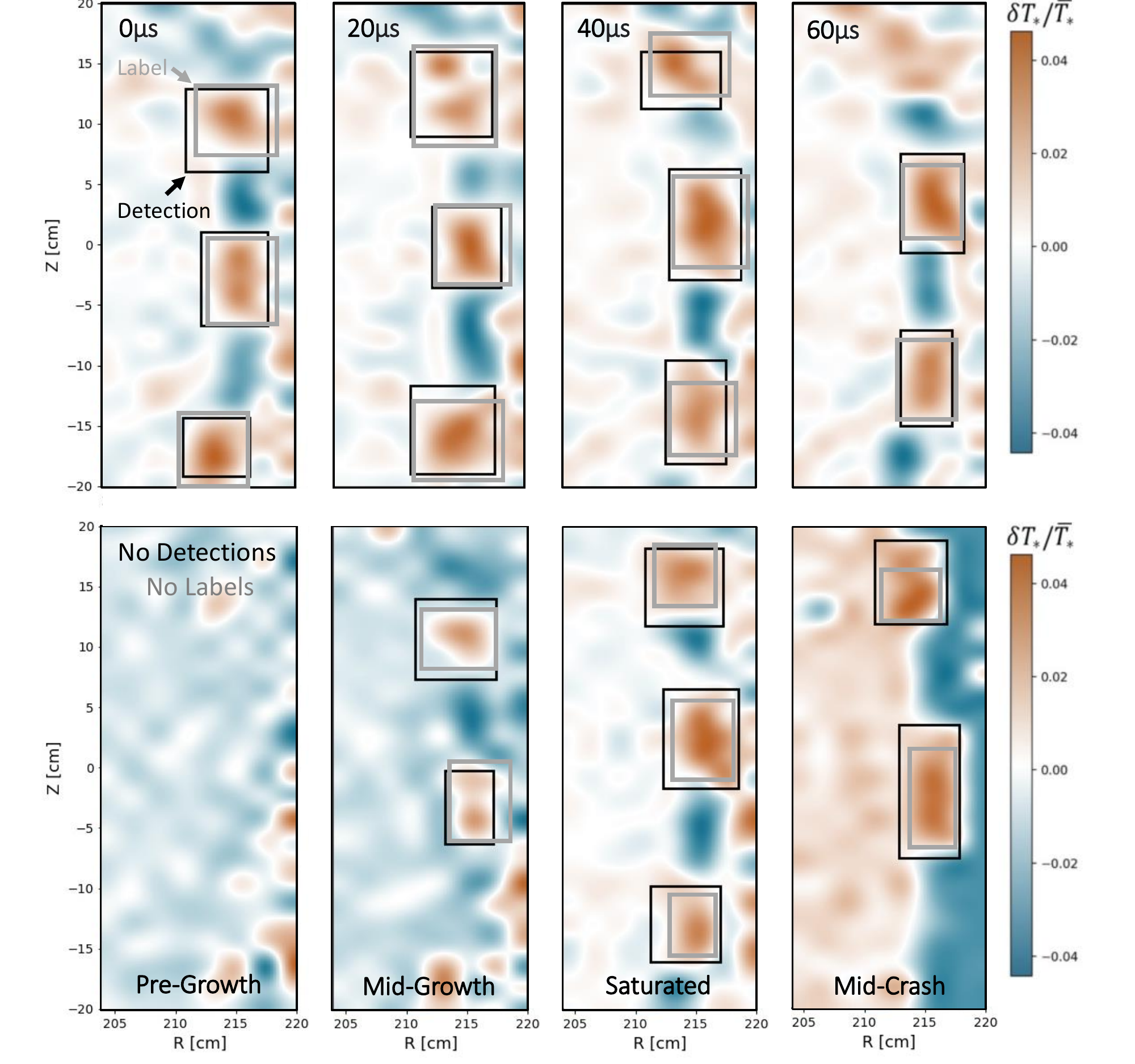}
  \caption{Examples of ELM filament detection in ECEI images, performed by a trained YOLO model. The upper row corresponds to the
  images in figure \ref{fig:ECEI_Sample} and the lower row corresponds to images during various stages in an ELM crash. Gray boxes 
  denote  manually labelled objects and black boxes denote predictions from the YOLO model. (This figure is an illustration, there is
  usually much more time between training images.)}
  \label{fig:detects}
\end{figure}

Figure \ref{fig:detects} presents ELM filaments detected by the trained YOLO model. The upper row shows ECEI data frames shown in
\ref{fig:ECEI_Sample}, but up-sampled to 192x96 pixels. The images shown in the lower row are taken from various phases of an
observed ELM crash. The gray boxes denote manually labeled bounding boxes and predictions from the trained YOLO model are illustrated by
black boxes. Visually inspecting the predictions we find that the center of the bounding boxes and their spatial extent
corresponds well to the manual labels. During the saturated and the mid-crash phases, however, they appear to be more
misaligned. The video supplementing this article further demonstrates the developed filament detection and tracking
method applied to a $10\,\mathrm{ms}$ long time window.

\begin{figure}[h!tb]
  \centering
  \includegraphics[width=1.0\textwidth]{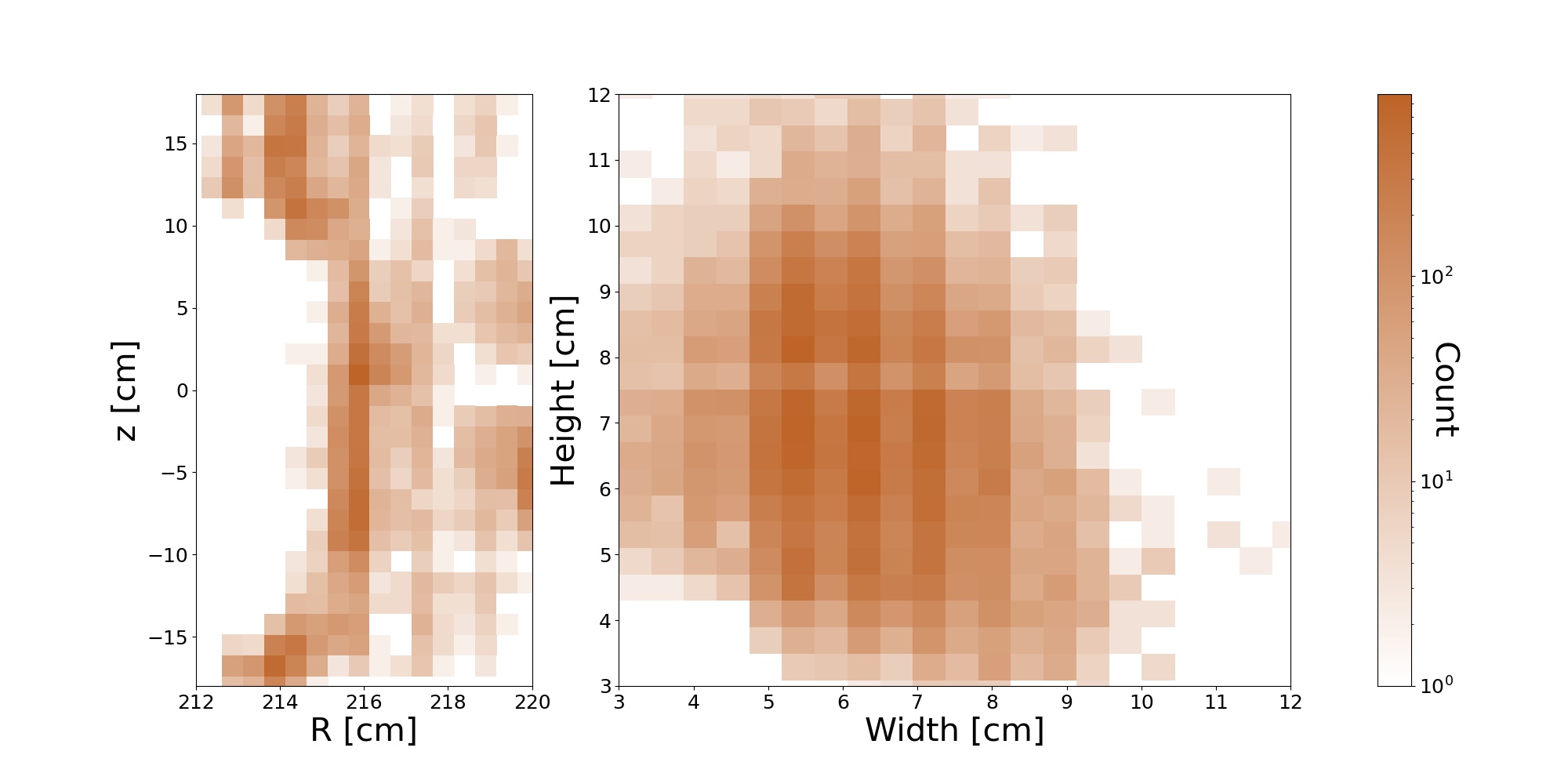}
  \caption{Histograms of the anchor and dimensions of the bounding boxes discovered by the trained machine learning model
  in the entire analyzed dataset.}
  \label{fig:det_dist}
\end{figure}

To investigate what degree of confidence is warranted in the predictions made by the trained YOLO model we plot histograms of the
spatial distribution of the detection boxes on unseen data in figure \ref{fig:det_dist} below. We find that the distribution of
anchor points now extends radially out over a single flux surface. Since CNNs are equivariant under translation this alone warrants no
loss of confidence in the predicted bounding boxes. But we furthermore observe that the distribution of bounding box dimensions exceeds
those prescribed in the training data. Detected bounding boxes of smaller width and height, between $3$ and $5\, \mathrm{cm}$ are 
frequently predicted but are absent in the training data. Since CNNs are not scale-invariant, these may be indicative of
lower model performance for tall and skinny or short and broad boxes. Examples of this can be seen in Fig.\ref{fig:detects} in the
mid-growth and mid-crash frames, where tall and skinny predicted bounding boxes appear to feature lower IoU than predicted
bounding boxes that appear more square-like.

Besides filament detection, the output of the trained YOLO model can be used to compile derived measurements, such as an average filament amplitude, ⟨A⟩.
\begin{align}
    \langle A \rangle = \frac{1}{N} \sum_{n=1}^{N} A_n \textrm{  where  } A_n = \sum_{i,j \in \mathrm{\textrm{bbox n}}} \left( \frac{\delta T}{\langle T \rangle}\right)_{i,j}^{+}. \label{eq:A}
\end{align}
This is approximately related to the average excess heat of all ELM filaments in a given ECEI frame. Here $i$ and $j$ index individual
image pixels within a given bounding box and $^{+}$ indicates that this calculation considers only positive-valued pixels, ignoring
negative-valued pixels sometimes found in the corners of the bounding boxes. 

Filament motion is tracked by connecting bounding boxes of filaments detected in subsequent frames with one another. In particular,
bounding boxes in consecutive frames with centers within 2cm of each other are taken to be the same filament. This method is applied to
a series of frames by first identifying filaments in all frames. Then, starting at the first frame,  identified filaments are associated with
filaments from successor frames. Then their radial and poloidal velocity is estimated using the center of the bounding box. If no successor
filaments can be identified, no velocity is calculated. We note here that the calculated filament velocities include contributions from background
electric drifts. While these can readily be removed by calculating the radial electric field from a force balance model, here we focus only on
properties of the filament tracking method developed. By detecting only peaks we do not assume any modal structure is present in the ECEI images.
This allows us to calculate filament amplitude statistics in all parts of the ELM crash development, from pre-growth to the final ELM crash. 

%%%%%%%%%%%%%%%%%%%%%%%%%%%%%%%%%%%%%%%%%%%%%%%%%%%%%%%%%%%%%%% Results %%%%%%%%%%%%%%%%%%%%%%%%%%%%%%%%%%%%%%%%%%%%%%%%%%%%%%%%%%%%%%%
\section{Data Analysis}
\label{sec:analysis}

\begin{figure}[h!t]
     \centering
     \begin{subfigure}[b]{0.45\textwidth}
         \centering
         \includegraphics[width=\textwidth]{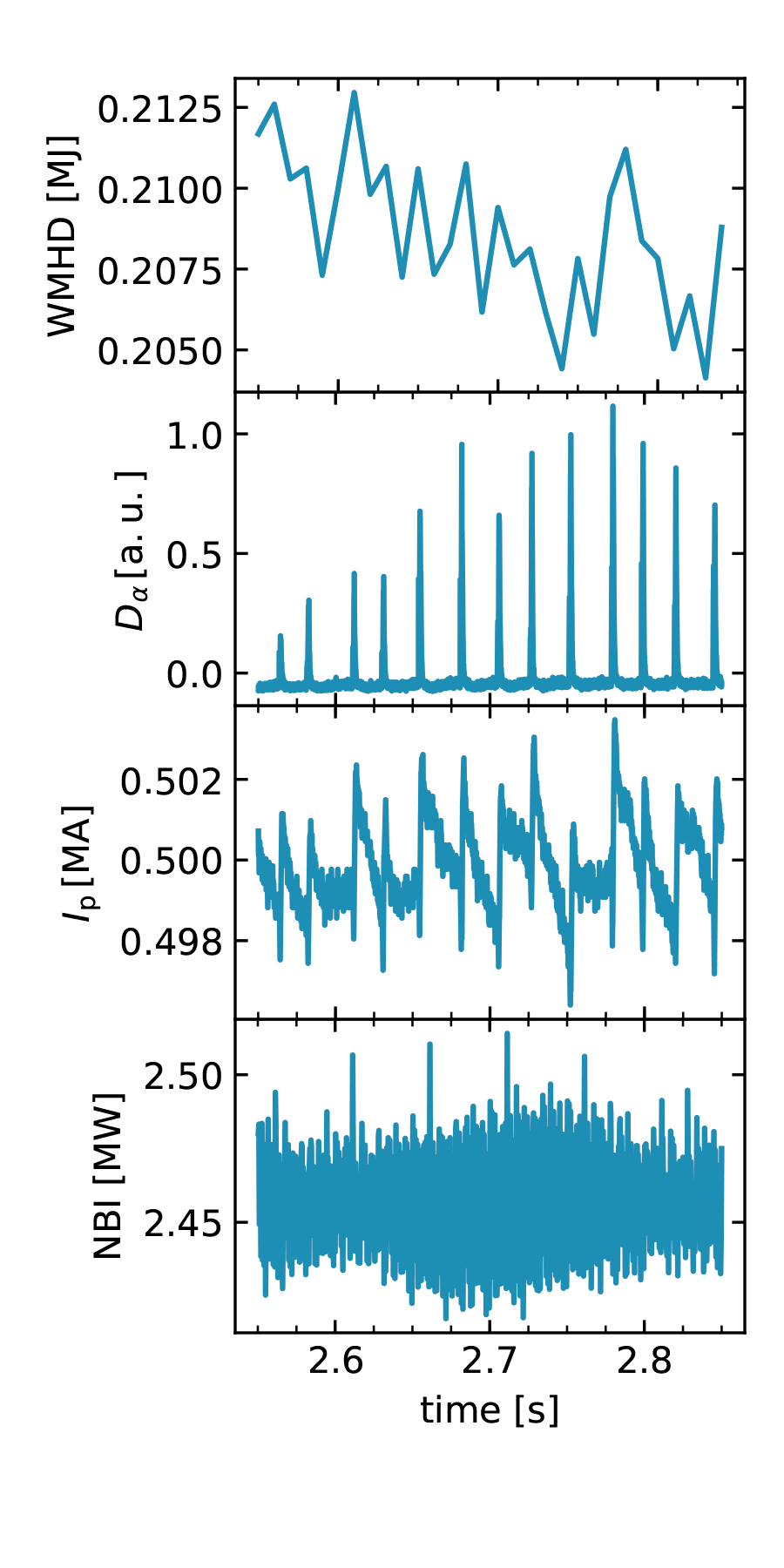}
         \caption{Time traces of global plasma quantities.}
         \label{fig:22289_timetrace}
     \end{subfigure}
     \hfill
     \begin{subfigure}[b]{0.4\textwidth}
         \centering
         \includegraphics[width=\textwidth]{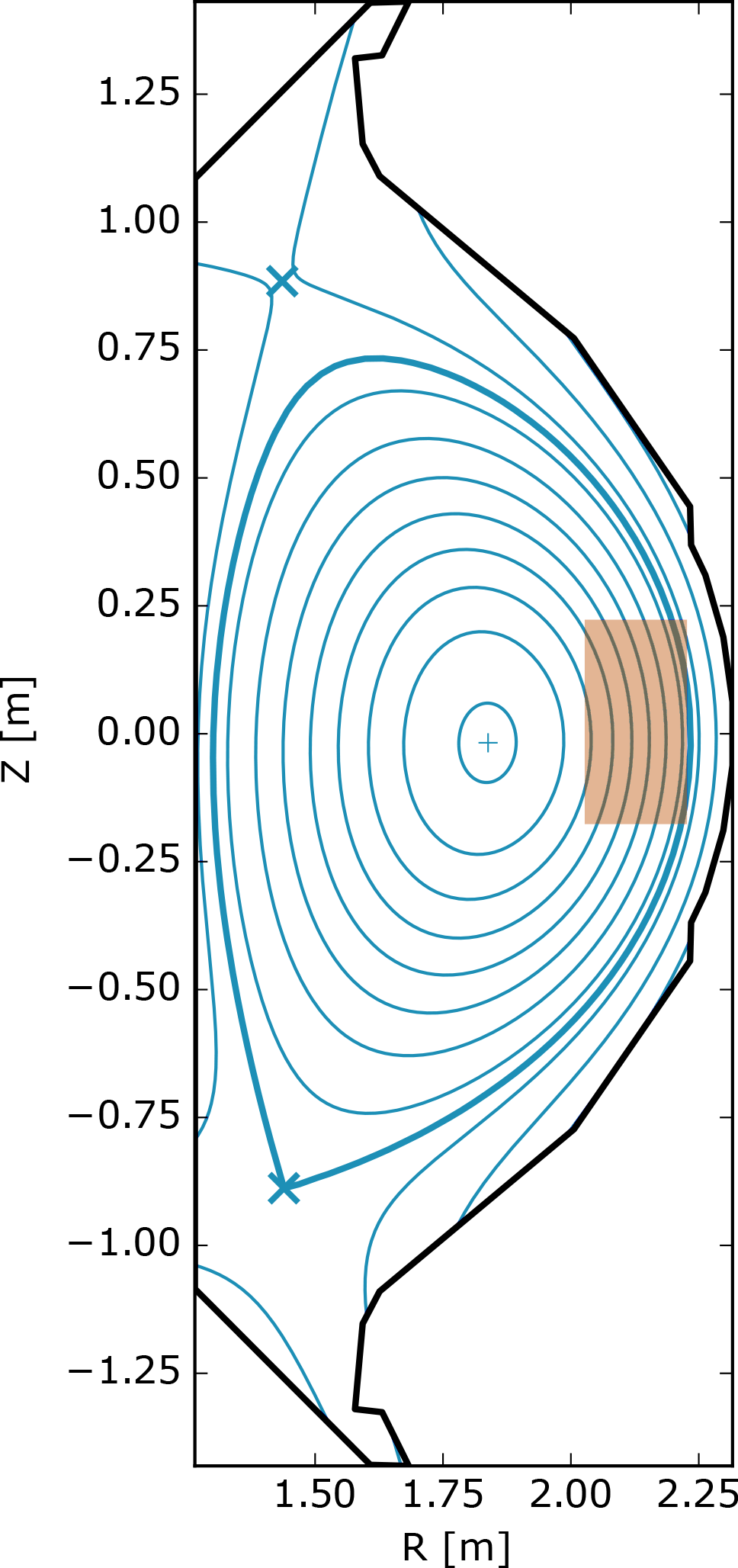}
         \caption{Magnetic equilibrium reconstruction, he orange rectangle denotes the view of the ECEI diagnostic.}
         \label{fig:22289_efit}
     \end{subfigure}
\caption{Global plasma state and magnetic equilibrium reconstruction for shot 22289}
\label{fig:22289_data}
\end{figure}

In the following, we use the trained YOLO model to investigate ELM filament dynamics for one H-mode plasma in KSTAR. This shot features a
toroidal magnetic field given by $B_\mathrm{T} = 1.8\mathrm{T}$ and was neutral beam heated. Figure \ref{fig:22289_timetrace} shows time traces
of the MHD energy, the emitted $D_\mathrm{\alpha}$ radiation, the plasma current, and the neutral beam power for shot 22289. The confined energy
decreases and sharp peaks of the measured $D_\alpha$ radiation indicate the presence of ELMs. The peaking frequency of $D_\alpha$ and the plasma
current $I_\mathrm{p}$ appear correlated while the neutral beam power remains approximately constant over the shown time slice. 
Figure \ref{fig:22289_efit} shows the magnetic equilibrium for this time slice, calculated using the EFIT \cite{lao-1985} module in OMFIT
\cite{omfit-2015}. The brown rectangle denotes the field of view of the ECEI diagnostic viewing the low-field side for which we report data 
analysis in the following.

\begin{figure}[h!tb]
  \centering
  \includegraphics[width=1.0\textwidth]{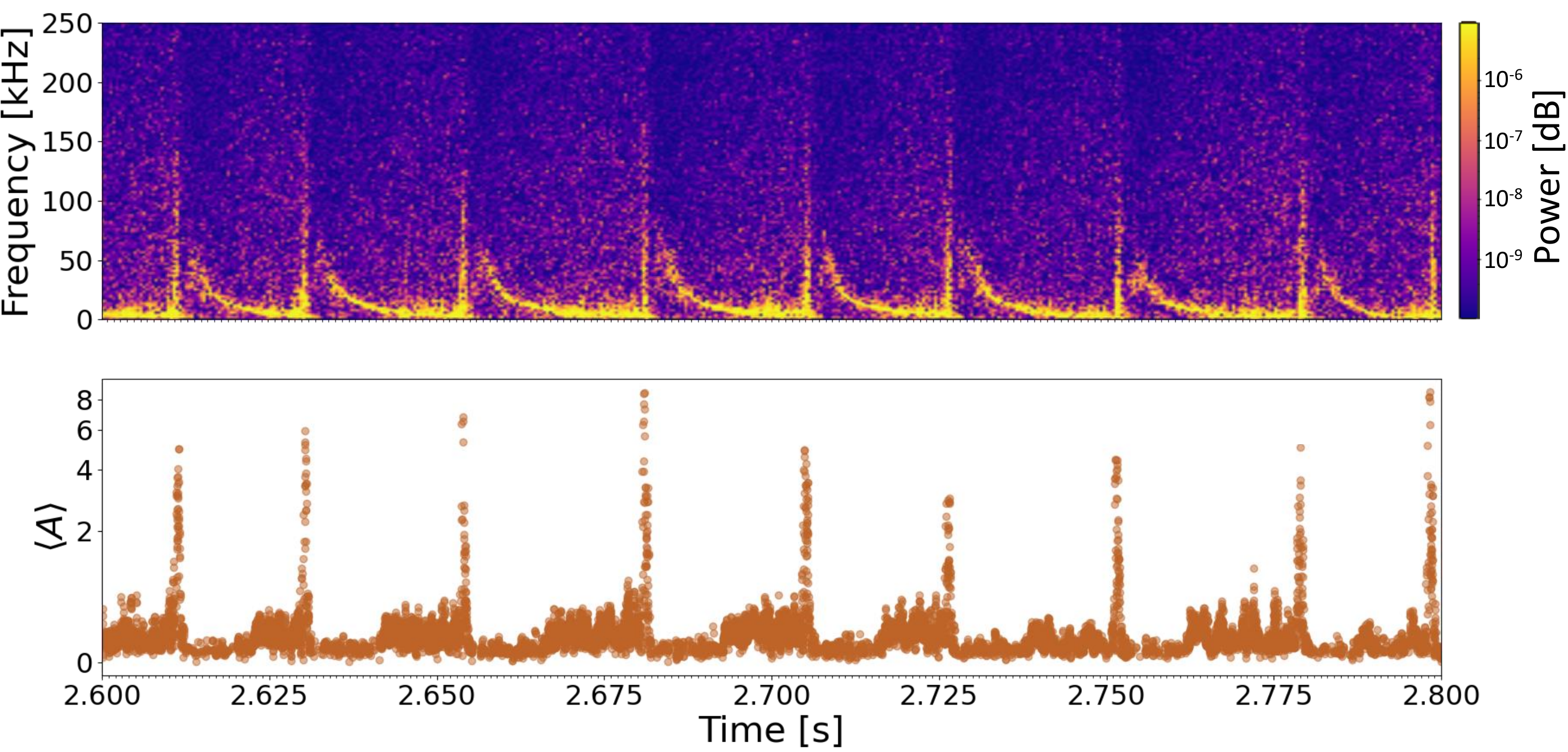}
  \caption{Spectrogram calculated using a single ECEi channel and evolution of \avgA.}
  \label{fig:22289_overview}
\end{figure}

As a first step, we investigate how the average filament amplitude \avgA, defined in  Eq.(\ref{eq:A}), behaves during ELM cycles. 
Figure \ref{fig:22289_overview} shows the time series of the spectrogram, calculated from ECEI data, together with the evolution of \avgA.
Here, ELM crashes appear as short bursts of broadband fluctuation in the spectrogram. In the inter-ELM crash period, the ELM mode
appears as a coherent fluctuation in the spectrogram and its frequency decreases with the pedestal recovery. Similarly, \avgA also
peaks during the ELM crash. The time evolution of \avgA, which is carried by detected filaments, can visually be divided into three states.
Immediately after an ELM crash, \avgA is approximately zero. The extent of this state appears to overlap with the frequency decreasing 
(pedestal recovery) period in the spectrogram in which the ELM amplitude may be too small to be detected clearly.  After this phase, 
\avgA exhibits a larger mean and fluctuations. This phase extends approximately over the same time intervals where only slow frequencies
below about $10\, \mathrm{kHz}$ appear in the spectrogram and precedes the ELM crash. During the actual ELM crash, \avgA sharply peaks to
values up to $5-10$, which is about two orders of magnitude larger than in the quiescent period that immediately follows the crash.

% We used our algorithm on multiple KSTAR shots other than 22289. It is worth noting that some other shots had no discernible modes 
% or multiple modes which interact or a number of other significant differences. 

\begin{figure}[h!tb]
    \centering
    \includegraphics[width=1.0\textwidth]{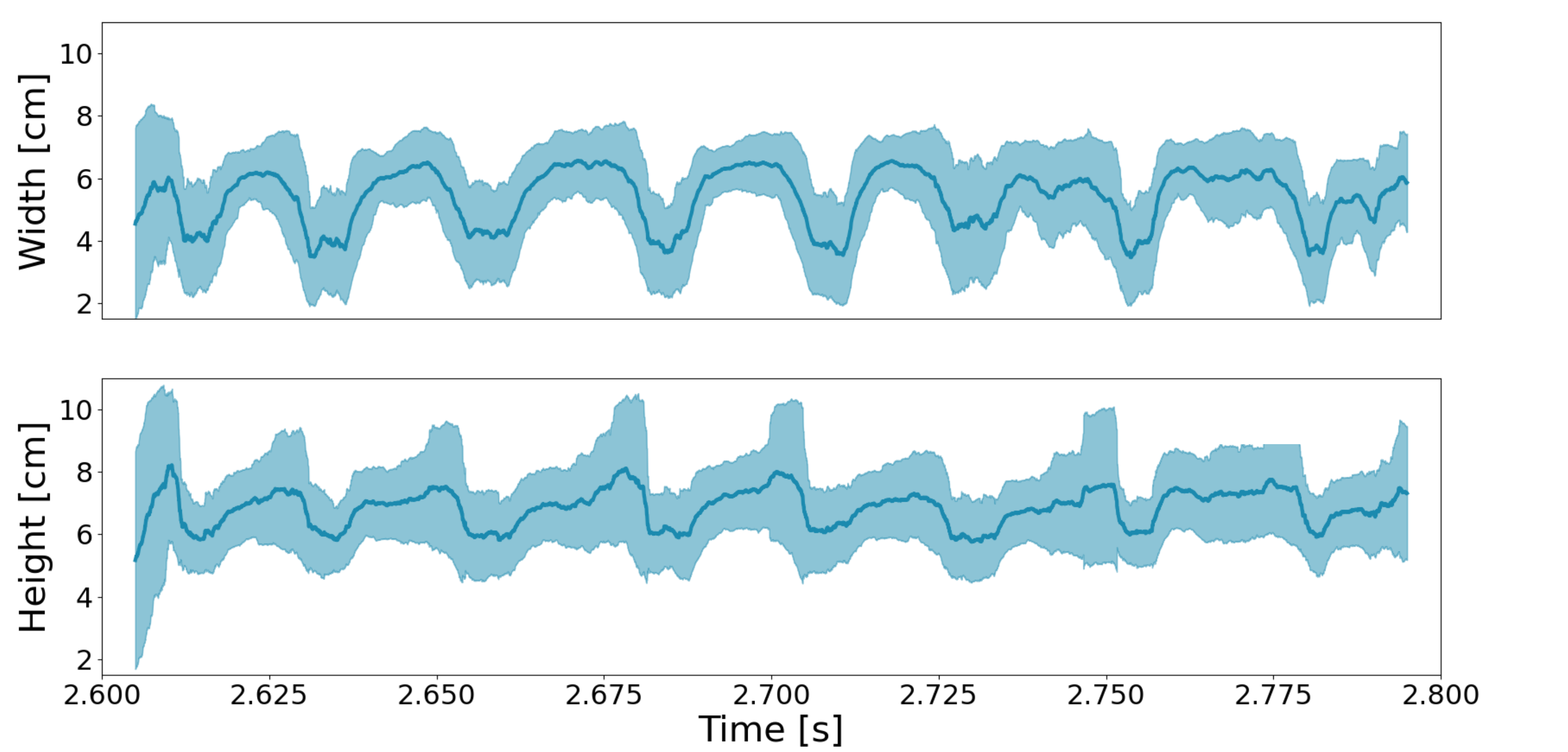}
    \caption{Width and height of detected filaments, calculated using a $0.01\,\mathrm{s}$ wide moving average window. The shaded
    area centered around the moving average denotes the moving standard deviation.}
    \label{fig:22289_detected_sizes}
\end{figure}

Figure \ref{fig:22289_detected_sizes} shows time-traces of the average bounding box width and height for the detected filaments,
calculated using a $10\, \mathrm{ms}$ long moving average window. The shaded area denotes the standard deviation, calculated
over the same interval as used for the moving average. Both, the average width and height undergo quasi-periodic oscillations
that are correlated with the ELM crashes shown in \ref{fig:22289_overview}. In the pre-crash phases, we observe average filament widths and heights
of approximately $6\, \mathrm{cm}$. About $1\, \mathrm{ms}$ before the ELM crash the average height of the mode increases abruptly while
its average width starts to decrease slowly, indicating that a significant change of the mode aspect ratio resulted in the ELM 
crash (see Fig.~\ref{fig:detects}). The appearance of the poloidally elongated mode just before the crash is consistent with the previous manual analysis
\cite{lee-2017} where it was identified as a solitary perturbation. After the ELM crash, the mode structures recover the 
pre-crash width and height. Figure \ref{fig:22289_detected_sizes} demonstrates that the machine-learning based model can be used to
identify ELM filament sizes for the temporal evolution of the ELM dynamics, extending through multiple pre-crash, growth, and crash phases.

\begin{figure}[h!tb]
    \centering
    \includegraphics[width=\textwidth]{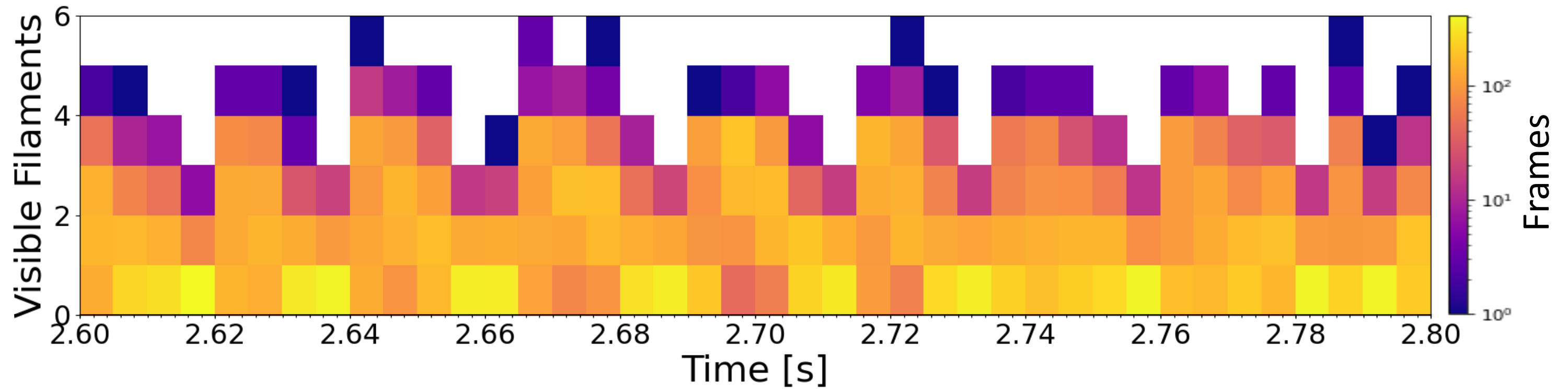}
    \caption{Histogram of detected filament count.}
    \label{fig:22289_detected_hist}
\end{figure}

The average number of filaments detected in each frame, shown in Fig.\ref{fig:22289_detected_hist}, shows
similar quasi-periodic oscillations. Here the individual histograms are calculated over intervals of $5\, \mathrm{ms}$.
For the pre-crash periods, the majority of frames feature zero
to three filaments, but single frames show up to 5 detected filaments. During the ELM crash, the number of
detected filaments decreases. Here, more frames are absent of any detected filaments and virtually no
frames in this period have more than three detected frames. This is possibly due to an increased poloidal wavelength. During the crash-phase, there are many frames where large regions of the image are hot (see Fig \ref{fig:detects}) or where the amplitude of the background turbulence is comparable to filaments, which often confuses the model.

\begin{figure}[h!tb]
  \centering
  \includegraphics[width=1.0\textwidth]{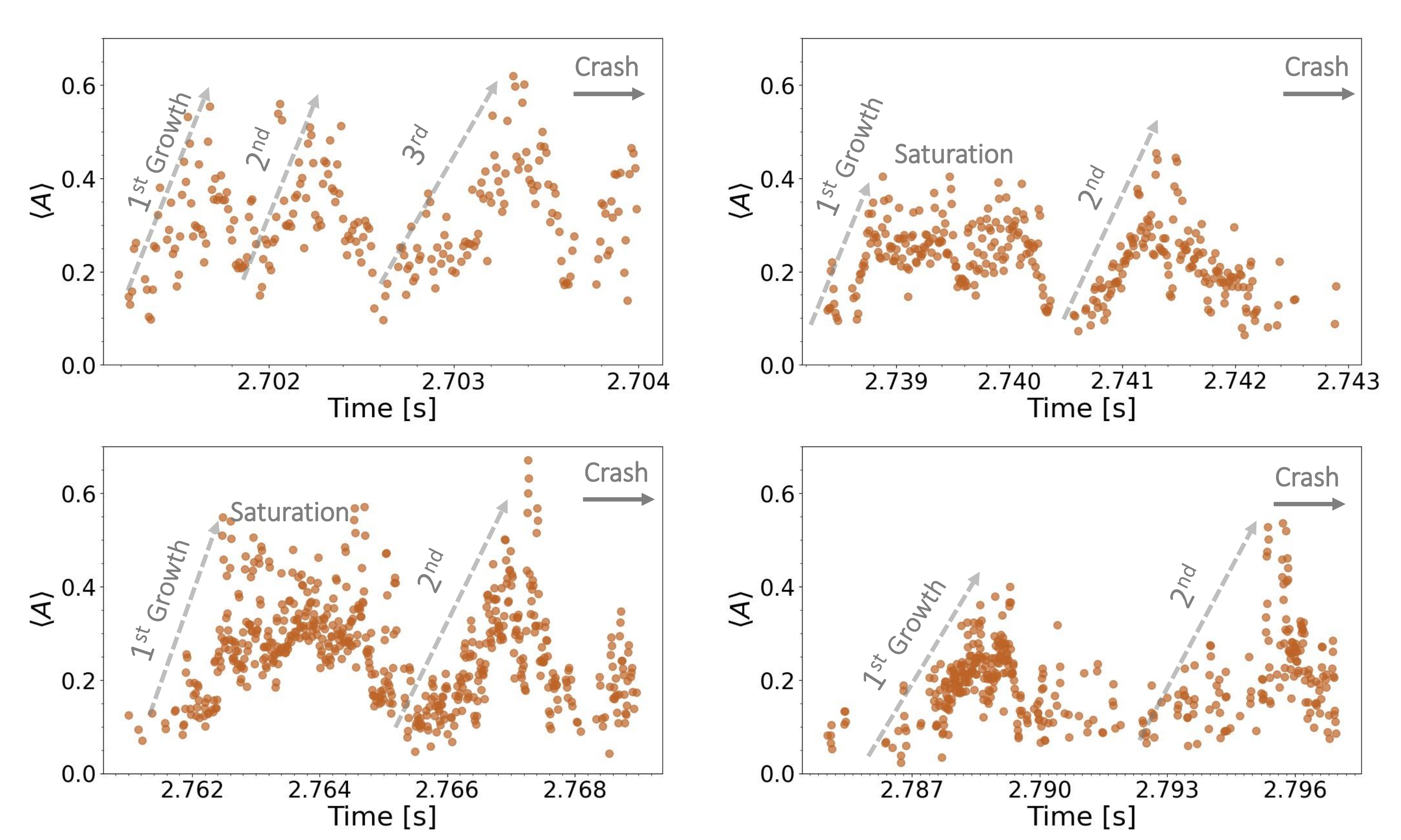}
  \caption{Amplitude dynamics of ELM filaments in the interval leading up to ELM crash.}
  \label{fig:22289_amplitude_detail}
\end{figure}

Figure \ref{fig:22289_amplitude_detail} shows the evolution of \avgA in a millisecond time interval right before ELM crashes. All four instances
show short growth periods where the total amount of heat contained in the ELM filaments increases by a factor of $2-4$. These growth periods may be
followed by a saturation period where \avgA varies only little. But we also observe that after a short growth phase, \avgA returns to its initial value.
Then \avgA may immediately begin to grow again, as observed for $t=2.701\,\mathrm{s}$ or remain at a low value before growing again, as observed at
$t=2.787\,\mathrm{s}$. The duration of the observed growth phases is about $1\, \mathrm{ms}$, regardless of whether \avgA saturates afterwards or not.

% While the amplitude experiences growth phases prior to many crashes in the previously shown shot, this is not true of every shot. Shots can also have crashes experiencing amplitude dynamics beforehand. An example of this behavior is shown in Figure \ref{fig:others_2} below (Shot 025522). 

% We also found examples of less-intense, miniature crashes which occurred just before a larger crash, without any other preceding amplitude dynamics. These pre-crash spikes appear in both the spectrogram and the plot of amplitude, as shown in Figure \ref{fig:others_2} below (Shot 025086). 

\begin{figure}[h!tb]
  \centering
  \includegraphics[width=1.0\textwidth]{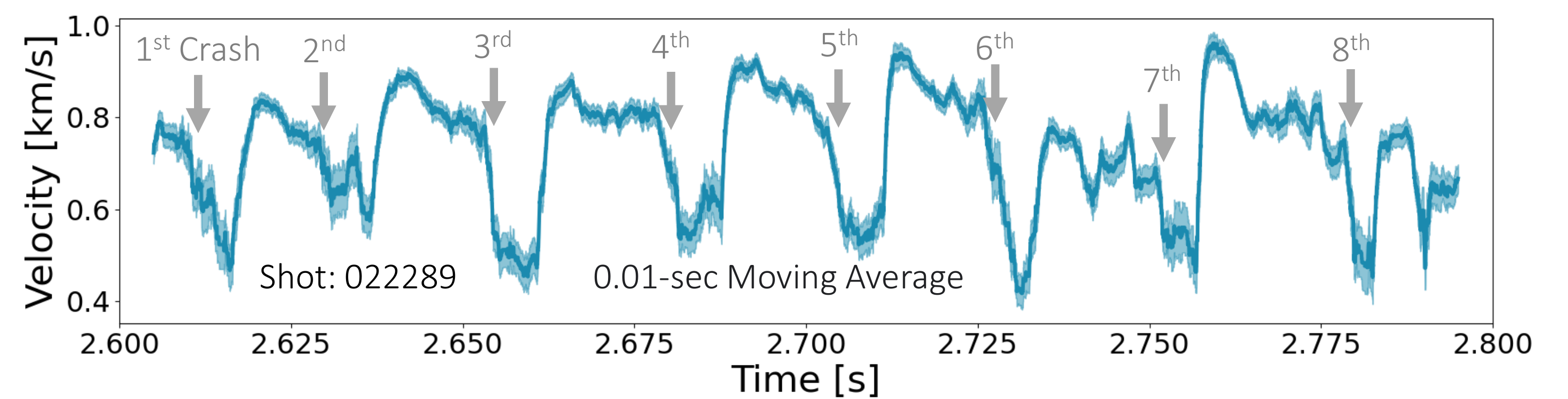}
  \caption{Moving average of poloidal filament velocity.}
  \label{fig:vel_comp_2}
\end{figure}

Figure \ref{fig:vel_comp_2} shows time series of the poloidal velocity.The ELM crash period is correlated with a sharp drop in poloidal filament velocity. 
In the intervals leading up to an ELM, for example between $t=2.615-2.63\,\mathrm{s}$ the poloidal velocity decreases slightly. While in other instances, 
for example during $t=2.66-2.68\mathrm{s}$, the poloidal velocity remains approximately constant. ELM crashes are associated with a rapid drop of the 
filaments poloidal velocity. In some instances, for example at $t=2.615\,\mathrm{s}$ the poloidal velocity immediately recovers to its pre-crash value while
in other instances, for example at $t=2.655\,\mathrm{s}$, the poloidal velocity remains attenuated for a short interval. 
%the poloidal filament velocity is about $0.8\, \mathrm{km/s}$. 
%This drops to
%approximately $0.4\, \mathrm{km/s}$ after an ELM crash. After dropping, the filaments velocity almost immediately recovers. We calculate an average poloidal velocity of $0.74 \pm 0.3 \mathrm{km/s}$. 

% By comparing the centers of each filament detections over time, the algorithm interprets consecutive detections as the same filament moving through space.  Using this tracking method, we estimate the average net poloidal flow along the pedestal region. For KSTAR shot 022289 we found this to be 0.74 km/s ±0.3. 

% By measuring this velocity over time, we find it has some relation with amplitude and that poloidal flow fluctuations often coincide with changes in the amplitude. The velocity tends to be is lower during the pre-saturation phases and accelerates by up to $\sim$0.45 km/s as a saturation phase begins. The velocity seems to undergo a small deceleration of up to $\sim$0.15 km/s leading up to a crash and a sudden, larger drop of up to $\sim$0.3 km/s during a crash (Fig. \ref{fig:vel_comp_2}). 

\begin{figure}[h!tb]
  \centering
  \includegraphics[width=0.6\textwidth]{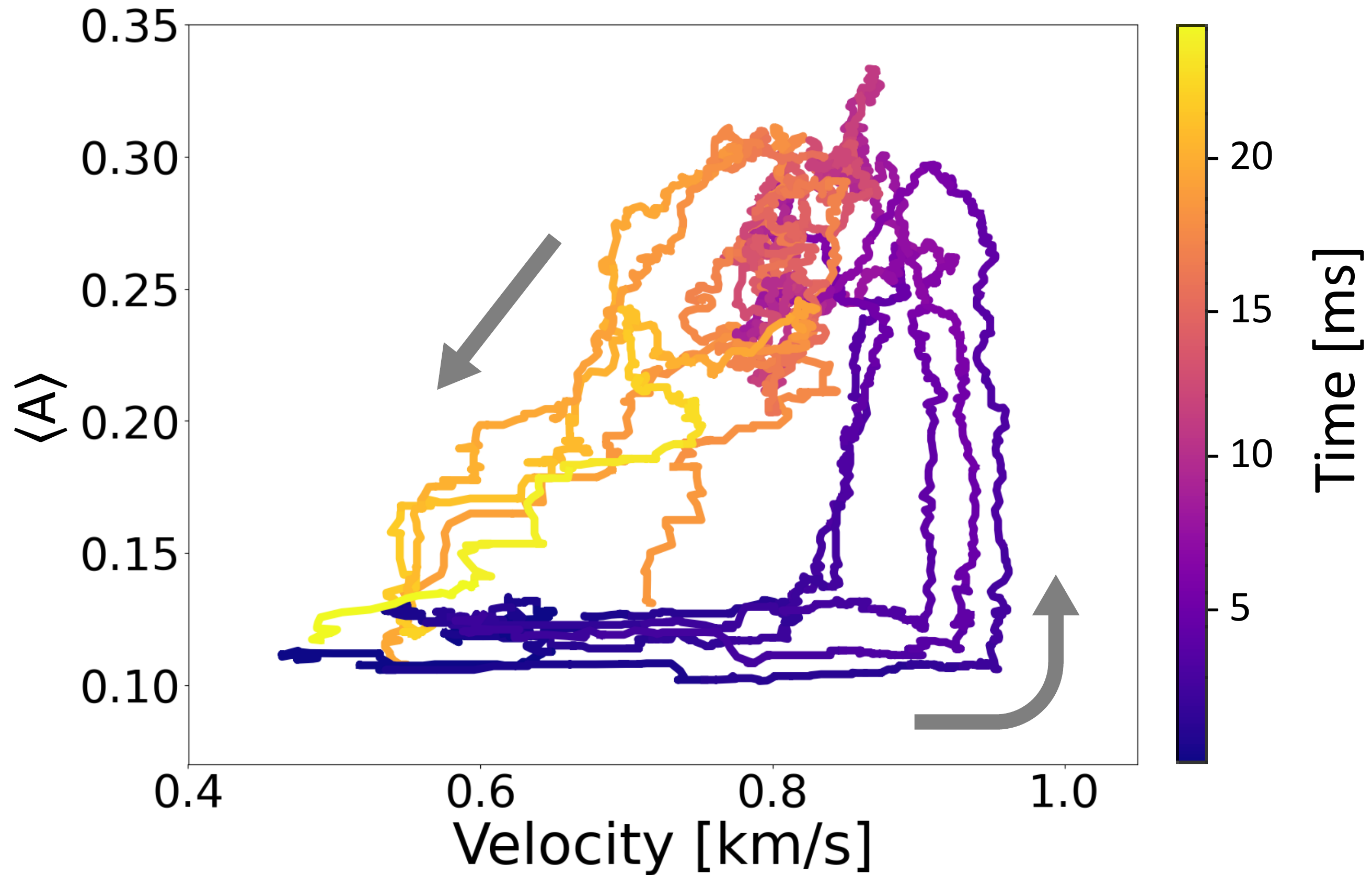}
  \caption{Trajectory of the ELM crash in the space spanned by average filament amplitude \avgA and poloidal velocity. 
  Time relative to an ELM crash is encoded in color. Five trajectories are overlaid.}
  \label{fig:amp_vel}
\end{figure}

Figure \ref{fig:amp_vel} visualizes the strong correlation cycle between filament amplitude and poloidal velocity. Tracked 
over 5 ELM crash cycles, we observe a hysteresis curve where first the poloidal velocity increases while \avgA remains constant. In the following 
pre-crash phase, the filament amplitude increases while the poloidal velocity remains constant. The ELM crash phase is then characterized by a
drop in filament amplitude from approximately \avgA $= 0.3$ to $0.1$, and a simultaneous drop in filament velocity from approximately 
$1.0 \mathrm{km/s}$ to $0.5 \mathrm{km/s}$. 

%%%%%%%%%%%%%%%%%%%%%%%%%%%%%%%%%%%%%%%%%%%%%%%%%%%%%%%%%%%%%%% Discussion %%%%%%%%%%%%%%%%%%%%%%%%%%%%%%%%%%%%%%%%%%%%%%%%%%%%%%%%%%%%%%%

\section{Discussion}\label{sec:discussion}
The YOLO-based detector reliably identifies ELM filaments in normalized ECEI images. As a supervised 
algorithm, the detector requires manually labeled training data to be able to detect ELM filaments in 
images. The trained detector achieves $93.7\%$ accuracy on a validation set and manual inspection of predicted ELM 
filament bounding boxes shows a good performance across various stages of ELM phase. The size distribution of the manually labeled 
data is narrower than the distribution of bounding boxes detected during inference. A manual inspection of the identified bounding
boxes suggests that the mismatching
part of the distribution is characterized by lower IoU scores. But regardless, the predicted bounding
boxes still reliably identify ELM filaments. This effect highlights the sensitivity of supervised
machine learning models to the biases included in the training data. In general, one needs to take
care of imbalance issues when developing data-driven methods and carefully compare the properties
of the training data set and the data set identified during inference. 

The trained detector is used to identify individual ELM filament crests for a $0.2\, \mathrm{s}$ long sub-interval
of an ELM cycle. From the detection events, we calculate the average width and height, filament amplitude, as well
as poloidal velocity. The bounding box dimensions of the detected filaments, as well as the count of detected filaments
present quasi-periodic behavior that is strongly correlated with the spectrogram as well as \avgA. Identifying the change 
of the bounding box as the mode-number of the ELM, this behavior is indicative of non-linear interactions which modify
the wavelength of the temperature perturbations. In particular, the width of the detected filaments changes from
approximately $6\, \mathrm{cm}$ before ELM crashes to approximately $4\, \mathrm{cm}$ during the ELM crash. That is, by a 
factor of 1.5. At the same time, far fewer but hotter filaments are detected on average during the ELM crash. 

A detailed analysis of \avgA in the periods leading up to an ELM crash shows a range
of behaviors. Small growth phases can be followed by an intermediate saturation and reversals.
This dithering between a ground state and a small elevated level appears random and no dynamics that
are predictive of a following ELM crash are observed in the analyzed data.

Finally, we find a strong correlation between \avgA and the poloidal filament velocity.
During each ELM crash, the average rotation velocity decreases by a factor of 2 from approximately $0.8\, \mathrm{km/s}$ to
approximately $0.4\, \mathrm{km/s}$. After each ELM crash the poloidal velocity rapidly recovers to pre-crash
levels. In the data shown here, there appears no characteristic time scale for any decrease
or increase in poloidal rotation velocity. For example, the velocity decreases gradually in the first or 6th ELM
crash shown in Fig.~\ref{fig:vel_comp_2} while the velocity drop is more spontaneous for the 3rd and 4th crash.
Finally, we show that the ELM cycle as observed in this single plasma follows a cyclical dynamic in the state space
spanned by \avgA and the poloidal velocity.

% We have implemented a machine-learning approach to enable the study of ELM filaments, as other researchers have recently begun doing for other sensors \cite{lee-2021}. By detecting individual ELM filaments on a frame-by-frame basis, the machine-learning-based detector allows us to infer the filament velocity also on a frame-by-frame basis. This is a major improvement over previous ECEI methods, where only average mode velocities are manually estimated from data, as described for example in \cite{vanovac-2018}. The observed velocities of about $0.5 - 1.0\mathrm{km/s}$ are similar to poloidal velocities reported at ASEDX upgrade \cite{boom-2011, vanovac-2018} and previous KSTAR analysis \cite{choi-2011}. We note here that the reported references are relative to the lab frame. Following an ELM crash, all measured properties revert to the values observed before the crash. We don't observe random changes in mode number \cite{lee-2017} which have been reported for KSTAR plasmas . In \cite{yun-2011} and \cite{Bogomolov_2015}, it was observed that the ELM filament pattern velocity decreases right before an ELM crash. Here we verify this observation and present additional statistics on this observation. Figure \ref{fig:vel_comp_2} shows that poloidal filament velocities can decrease by about $0.1\mathrm{m}/\mathrm{s}$ right before an ELM crash (The exact value is hard to calculate due to the fluid boundaries of the ELM phases). We suggest that the rapid transition shown in figure \ref{fig:amp_vel} may be related to the model suggested in \cite{oh-2018}. 

The measurements of our machine-learning detector compare favorably with previous manual analyses, for example the reported
ELM filament velocity range of 0.5 to 1.0 km/s agrees well with results presented for both, 
KSTAR \cite{choi-2011, vanovac-2018, choi-2011, lee-2017} and ASDEX upgrade\cite{boom-2011, vanovac-2018}.
Both \cite{yun-2011} and \cite{Bogomolov-2015} report that ELM filament velocities decrease immediately
before an ELM crash. Here, we verify this observation, and discover a hysteresis relation between the velocity
and the amplitude. It has previously been suggested that radio frequency wave emission \cite{M-Kim_2020} might be responsible
for this velocity reduction before the ELM crash and the hysteresis relation. Also, we found that the mode aspect ratio changes
significantly just before the ELM crash, which is consistent with previous observations of solitary perturbation \cite{lee-2017}.

%%%%%%%%%%%%%%%%%%%%%%%%%%%%%%%%%%%%%%%%%%%%%%%%%%%%%%%%%%%%%%% Conclusion %%%%%%%%%%%%%%%%%%%%%%%%%%%%%%%%%%%%%%%%%%%%%%%%%%%%%%%%%%%%%%%
\section{Summary and outlook}
\label{sec:conclusion}
We use a machine learning model, based on the YOLO-v4 classifier, to detect ELM filaments in ECEI images. The developed
detector performs robustly and is used to identify bounding boxes of ELM filaments during a $0.2\,\mathrm{s}$ long sub-interval.
This data is used to investigate ELM filament dynamics. In particular, we compile filament dimensions, the average ELM filament amplitude,
as well as the poloidal velocity of the filaments in the laboratory frame. For the analyzed ELM crashes, all these quantities present
quasi-periodic behavior. Right after an ELM crash, the average filament amplitude is low, the average filament is about
$4\, \mathrm{cm}$ wide and $6\, \mathrm{cm}$ tall and rotates counter-clockwise with about $0.8\, \mathrm{km/s}$. Leading up
to an ELM crash, the average filament amplitude increases, and their width, as well as poloidal velocity, decreases. The average number 
of filaments visible per frame also decreases in the period leading up to an ELM crash. In other words, ELM crashes manifest as few, hot
filaments. The quasi-periodic behavior can be interpreted as closed cycles in the phase space spanned by \avgA and the poloidal
velocity.

Future work will focus on systematic investigations of ECEI data sampled during ELM cycles. The possibility to automatically collect
variations in filament properties on a frame-by-frame basis may lead to further discovery. For example, this may aid in identifying
solitary perturbations and distinguishing them from filamentary ELM structures, this allowing a more targeted investigation into the
non-linear evolution of the underlying ELM instability physics.  And finally, it may be desirable to condense the convolutional
architecture of the used YOLO model into a simpler one, with still enough flexibility to identify ELM filaments in ECEI data.

% The trained detector is used to 
% In previous works, filament amplitude has been calculated similarly but for far fewer frames as filaments were labeled manually. Poloidal velocity along the pedestal region has also been estimated previously, but not with enough granular detail to ascertain any dynamics. The dynamics of the mode aspect ratio seems to be a new observation.

% We have used the detector to compile ELM filament statistics for H-mode plasmas. The studied plasma
% present quasi-periodic ELMs, as indicated by $D_\alpha$ measurements. We used our trained filament
% detector to compile linear sizes, integrated temperature, as well as poloidal velocity of the
% ELM filaments. The picture that emerges from this study is the following:
% All quantities feature strongly correlated variations over the entire ELM cycle.
% In the pre-crash period, a fast-rotating mode structure is observed. While during the ELM crash
% this poloidal rotation is damped and fewer filaments carry substantial heat away from the confined plasma.

% Future work will include a systematic study of ELM filament motion in H-mode plasmas where we aim to identify different regimes of ELM filament dynamics
% and connect them to the discharge parameters. We will also aim to simplify the used convolutional neural network architecture which may allow to
% use less training data while keeping detection accuracy constant. 

%%%%%%%%%%%%%%%%%%%%%%%%%%%%%%%%%%%%%%%%%%%%%%%%%%%%%%%%%%%%%%% Acknowledgements %%%%%%%%%%%%%%%%%%%%%%%%%%%%%%%%%%%%%%%%%%%%%%%%%%%%%%%%%%%%%%%
\section{Acknowledgements}\label{sec:Acknowledgements}
This work was made possible by funding from the Department of Energy for the Summer Undergraduate Laboratory Internship (SULI) program. This work is supported by the US DOE Contract No. DE-AC02-09CH11466 and R\&D Programs of ''KSTAR Experimental Collaboration and Fusion Plasma Research (EN2201-13)''. The authors report no conflicts of interest. 

\vspace{10pt}

\bibliographystyle{style/ans_js}                                                                           %custom ANS journal submission template bibliography style
\bibliography{bibliography}

\end{document}